\documentclass[a4paper,11pt]{article}
\pdfoutput=1 

\usepackage{jheppub} 

\usepackage[T1]{fontenc} 

\usepackage{comment}
\usepackage{graphicx,rotating,amsmath,amsfonts,amssymb}
\usepackage{colour}

\newcommand{\td}{\mathrm{d}}

\newcommand{\be}{\begin{equation}}
\newcommand{\ee}{\end{equation}}
\newcommand{\bea}{\begin{eqnarray}}
\newcommand{\eea}{\end{eqnarray}}
\newcommand{\ba}{\begin{array}}
\newcommand{\ea}{\end{array}}
\newcommand{\nn}{\nonumber}
\newcommand{\bi}{\begin{itemize}}
\newcommand{\ei}{\end{itemize}}
\newcommand{\ben}{\begin{enumerate}}
\newcommand{\een}{\end{enumerate}}
\newcommand{\bc}{\begin{center}}
\newcommand{\ec}{\end{center}}

\newcommand{\PRexp}{2.14 \pm 0.05}

\newcommand{\beq}{\begin{equation}}
\newcommand{\eeq}{\end{equation}}

\newcommand{\gsim}{\lower.7ex\hbox{$\;\stackrel{\textstyle>}{\sim}\;$}}
\newcommand{\lsim}{\lower.7ex\hbox{$\;\stackrel{\textstyle<}{\sim}\;$}}

\newcommand{\MeV}{{\rm MeV}}
\newcommand{\TeV}{{\rm TeV}}
\newcommand{\GeV}{{\rm GeV}}
\newcommand{\fb}{{\rm fb}}
\newcommand{\pb}{{\rm pb}}

\newcommand{\Om}{\Omega}

\newcommand{\al}{\alpha}
\newcommand{\bt}{\beta}
\newcommand{\ga}{\gamma}

\newcommand{\la}{\lambda}

\newcommand{\sa}{\sigma}

\newcommand{\Ga}{\Gamma}

\newcommand{\La}{\Lambda}

\newcommand{\cL}{\mathcal{L}}

\newcommand{\rar}{\rightarrow}

\newcommand{\dof}{N_{\rm dof}}

\title{Non-minimal CW  inflation, electroweak symmetry breaking and the 750 GeV anomaly 
}

\author[1,2]{L. Marzola,}
\author[1]{A. Racioppi,}
\author[1,2]{M. Raidal,}
\author[1]{F. R. Urban,}
\author[1]{H. Veerm\"ae}
\affiliation[1]{National Institute of Chemical Physics and Biophysics, R\"avala 10, 10143 Tallinn, Estonia.}
\affiliation[2]{Laboratory of Theoretical Physics, Institute of Physics, University of Tartu; Ravila 14c, 50411 Tartu, Estonia.}

\abstract{
We study whether the hinted 750 GeV resonance at the LHC can be a Coleman-Weinberg inflaton which is non-minimally coupled to gravity. Since the inflaton must couple to new charged and coloured states to reproduce the LHC diphoton signature, the same interaction can generate its effective potential and trigger the electroweak symmetry breaking
via the portal coupling to the Higgs boson. This inflationary scenario predicts a lower bound on the tensor-to-scalar ratio of $r\gtrsim 0.006$, where the minimal value corresponds to the measured spectral index $n_s\simeq0.97$. However, we find that the compatibility with the LHC diphoton signal requires exotic new physics at energy scales
accessible at the LHC. We study and quantify the properties of the predicted exotic particles.}

\begin{document}

\maketitle

\section{Introduction}
\label{sec:in}

It is widely believed that our Universe kicked off with a bang: an epoch of rapid accelerated expansion --- known as inflation~\cite{Starobinsky:1980te,Guth:1980zm,Linde:1981mu} --- driven by the potential energy of the inflaton field.  The inflaton typically sets out in an unstable region of its potential and, in the most common scenarios, very slowly rolls down towards a minimum. As it rolls, the field acquires an increasingly larger  kinetic energy which, eventually, dominates over the potential and puts an end to the accelerated expansion era. The inflaton then decays and reheats the stretched and empty causally connected patches of spacetime it generated, producing the primordial soup which has been cooling down until today.  One of the most remarkable achievements of this simple framework is the generation of the seeds of the primordial spacetime perturbations which evolve into all the structures we observe today, from the linear Cosmic Microwave Background (CMB) fluctuations, to Galaxies, Clusters, and beyond~\cite{Ade:2015xua,Ade:2015ava,Ade:2015lrj}.

In the Standard Model (SM) of particle physics there is no obvious candidate to play the r\^ole of the inflaton, which is often modelled after some new scalar field for simplicity.  The only scalar in the SM, the Higgs field, unfortunately has interactions that are far too strong to result in the shallow potential required by inflation. In other words, the Higgs potential is so steep that the friction provided by the expansion of the Universe is not enough to keep the field in the slow-roll regime that inflates our Universe. However, the pace at which the Higgs field rolls down its potential  can be controlled by postulating new non-minimal interactions between the field and gravity, which have the effect of flattening the potential at very large values of the field itself. The resulting ``Higgs inflation'' model~\cite{Bezrukov:2007ep} reproduces the observed Universe quite successfully; one of the virtues of this construction --- or perhaps one of its vices --- is that nearly everything is strictly determined by our knowledge of particle physics and cosmological observations. For instance, given the now measured values of the Higgs boson and top quark masses, and in absence of some new phenomena below the Planck scale, the model predicts that our Universe has been developing in a metastable state~\cite{Buttazzo:2013uya}.

Fortunately, there may be another possibility. Recently, the ATLAS~\cite{ATLAS2015} and the CMS~\cite{CMS2015} collaborations reported an excess of events in the diphoton channel at an invariant mass of about 750~GeV. The two signals seem compatible and could be generated by the decay of a new spin zero or spin two particle. Given the production cross section times diphoton branching ratio of about $\sigma(pp\rar\phi)\rm{Br}(\phi\rar\ga\ga)\approx 6$~fb \cite{DiChiara:2015vdm} and the absence of signals in complementary channels as the di-jet~\cite{Khachatryan:2015dcf} and the $t\bar t$ one~\cite{Khachatryan:2015sma}, we ascribe this anomaly to a new scalar particle which couples to the SM gauge bosons via scalar mediators~\cite{Gabrielli:2015dhk,Knapen:2015dap}. As we will show below, reproducing the LHC diphoton signal requires that the effective couplings of the new particles to the SM are too large for it to act as an inflaton itself. However, once we allow for a non-minimal coupling between the new particle and gravity, we are able to recover a viable and predictive inflationary scenario in which all the masses of the theory, with the exclusion of the scale related to gravity, are generated radiatively via the Coleman-Weinberg (CW) mechanism~\cite{Coleman:1973jx}.

On the phenomenological side, the outcomes of our construction depend on the value of the portal coupling between the mediators and the inflaton and on the vacuum expectation value (vev) of the latter, yielding a minimalistic scenario that is accessible by future collider experiments and cosmological observations. As shown in~\cite{Kadastik:2011aa},
for the measured Higgs boson mass of 125~GeV, a small portal coupling of the Higgs boson to a scalar singlet (the inflaton in our case) provides a good solution to the SM vacuum stability problem. The resulting scale invariant SM~\cite{Farzinnia:2013pga,Gabrielli:2013hma} remains perturbative up to scales exceeding the Planck scale and the electroweak scale is
generated through the Higgs portal coupling. However, for this to be the case, it is crucial that the scalar self- and portal couplings be smaller than unity.
As we will show below, the large cross section of the LHC 750~GeV anomaly challenges this requirement and introduces a tension between the cosmological scenario and the LHC diphoton signal.

In this work we detail the proposed non-minimal CW inflation and study its compatibility with the hinted 750~GeV anomaly at the LHC.
We find that this inflationary scenario is predictive and cosmologically viable. Because of the predicted tensor-to-scalar ratio $r\gsim 0.006,$ for the measured
value of the scalar spectral index $n_s\approx 0.97$ our model can be distinguished from other presently favoured inflationary scenarios such
as linear inflation~\cite{Kannike:2015kda}, Higgs inflation~\cite{Bezrukov:2007ep} and Starobinsky inflation~\cite{Starobinsky:1980te}.
Whereas the possibility of identifying the hinted LHC resonance with the non-minimally coupled inflaton was recently mentioned in~\cite{Dhuria:2015ufo,Salvio:2015jgu,Hamada:2015skp},
we now present a complete picture of this scenario within a particularly well motivated framework.
The outcome of our study is that such identification is rather constrained because the LHC diphoton signal typically requires very large couplings in the scalar sector which, in turn, result in Landau poles at scales not far above the TeV. As a consequence the predictions of the scenario become unphysical already at scales much below the inflation one. To avoid this problem, we find that the number of new degrees of freedom that connect our inflaton to the SM must be large. As we will show, the scalar mediators must then belong to large representations of the SM or, alternatively, must have unusually large electric charges and/or large dimension-full trilinear couplings to the CW inflaton itself. We conclude that in order to identify the 750~GeV resonance with the latter, a large sector of exotic particles must be
accessible at the LHC.

The paper is organised as follows: in Sec.~\ref{sec:cwi} we introduce our model of CW inflation and discuss the details of the reheating process in Sec.~\ref{ssec:reh}. In Sec.~\ref{sec:dip} we discuss the collider phenomenology of the model in strict connection to the mentioned LHC diphoton excess, while in Sec.~\ref{sec:end} we draw our conclusions.

\section{A new Coleman-Weinberg inflation scenario}
\label{sec:cwi}

We take a classically scale invariant matter sector non-minimally coupled to gravity:
\be
  \cL_\phi = - \frac{\xi \phi^2+M^2}{2}R - \frac{(\nabla\phi)^2}{2} - V(\phi) \, ,
  \label{eq:fullL}
\ee
where $\phi$ is the inflaton, non-minimally coupled to the Ricci scalar $R$ through the coupling $\xi$.  The effective Planck mass is $\xi \phi^2 + M^2 = M_\text{Pl}$, with $M$ being the bare Planck mass and $M_\text{Pl}=2.4\times10^{18}$~GeV. We set $M \approx M_\text{Pl}$ because, as we will show below, the vacuum expectation value of the inflaton $v_\phi$ is much smaller than $M_\text{Pl}$ itself.

We write the generic 1-loop scalar potential $V(\phi)$ generated from a classically scale invariant potential as
\be
  V(\phi) = \frac14 \la_\phi(\phi)\phi^4 = \frac18 \bt_{\la_\phi} \left( \ln \frac{\phi^2}{v_\phi^2} - \frac12 \right) \phi^4 \, ,
  \label{eq:Veff}
\ee
where $\bt_{\la_\phi}$ is the beta function of the inflaton self-coupling $\la_\phi$.
The potential $V(\phi)$ has been projected onto the direction where inflation happens, that is where any other scalar field is frozen at the minimum of its potential.   As is customary within inflation model building, we add to the bare potential, and leave henceforth understood, since numerically irrelevant, a bare cosmological constant term $\La^4$, which is tuned in order to achieve the present vanishing vacuum energy $V(v_\phi)=0$.

In order to explain the LHC data, we introduce one or more further scalar fields with non-vanishing electric charge and colour: $\sigma$. On top of the corresponding kinetic terms we then expand our classically scale-invariant Lagrangian with the following interactions\footnote{The quartic couplings of the scalar mediators $\sigma$ generally have a more complicated structure for larger representations and inclusion of many generations, e.g. for multiple generations of mediators in the $SU(3)_{c}$ fundamental we should include the term $\lambda^\prime_\sigma \left|(\sigma_i^\dagger\sigma_j)\right|^2$. These terms will, however, not change our conclusions at NLO and are thus omitted for the sake of readability.}
\be
	V(\phi\sa) = \frac14 \la_\phi \phi^4 + \frac12 \la_{\phi\sa} \phi^2 |\sa|^2 + \la_\sa |\sa|^4 \, , \label{eq:Vs_col}
\ee
which result in a beta function
\be
	\bt_{\la_\phi} \simeq \dof \left(\frac{\la_{\phi\sa}}{4\pi}\right)^{2} \, , \label{eq:beta_colour}
\ee
where $\dof$ denotes the number of degrees of freedom running in the loop. In a scenario with $N_g$ generations of the mediator particles in a $d_r$ dimensional colour representation $r$ we have $\dof = d_r N_g$. The $\beta$-function in Eq.~(\ref{eq:beta_colour}) is obtained under the assumption that the dominant contribution is due to the portal interactions. Moreover in the following we also assume that the running of the non-minimal coupling $\xi$ is negligible since the condition $\beta_\xi \ll \xi$ is realized. For more details about this approximation we refer the reader to~\cite{Kannike:2014mia,Kannike:2015apa,Kannike:2015kda}. We remark that our setup can also be extended to accommodate additional interactions between the SM Higgs field, the inflaton and the scalar mediators, which can be written in same fashion as in Eq.~\eqref{eq:Vs_col}. These new couplings result in the dynamical generation of the EW scale via the CW mechanism and solve the Higgs vacuum stability problem \cite{Gabrielli:2013hma}. In the following we will implicitly assume these features referring the reader to \cite{Gabrielli:2013hma} for further details. We also remark, that similar models of CW inflation were studied in \cite{Okada:2011en,Okada:2015lia}, involving a $U(1)_{B-L}$ gauge group and leading to results in line with ours.

The portal interaction included in Eq.~\eqref{eq:Vs_col} induces a mass for the inflaton
\be\label{eq:mphi}
  m^2_\phi = \bt_{\la_\phi} v_\phi^2 \, ,
\ee
while the mass of the mediator $\sa$ is
\be
  m^2_\sa = \frac{1}{2} \la_{\phi\sa} v_\phi^2 \, .
\ee
It follows, that for couplings in the perturbative regime the model can naturally accommodate a mediator particle $\sa$ heavier than the inflaton:
\be\label{eq:msigma}
  m_\sa = \frac{4\pi}{\sqrt{2\la_{\phi \sa}\dof}}  \, m_\phi.
\ee
However, it will be shown in the next section that the two masses need to be of the same order to explain the diphoton excess, yielding large or even non-perturbative couplings for contained scalar mediator sectors.

\begin{figure}[t!]
\bc
  \includegraphics[width=0.49\textwidth]{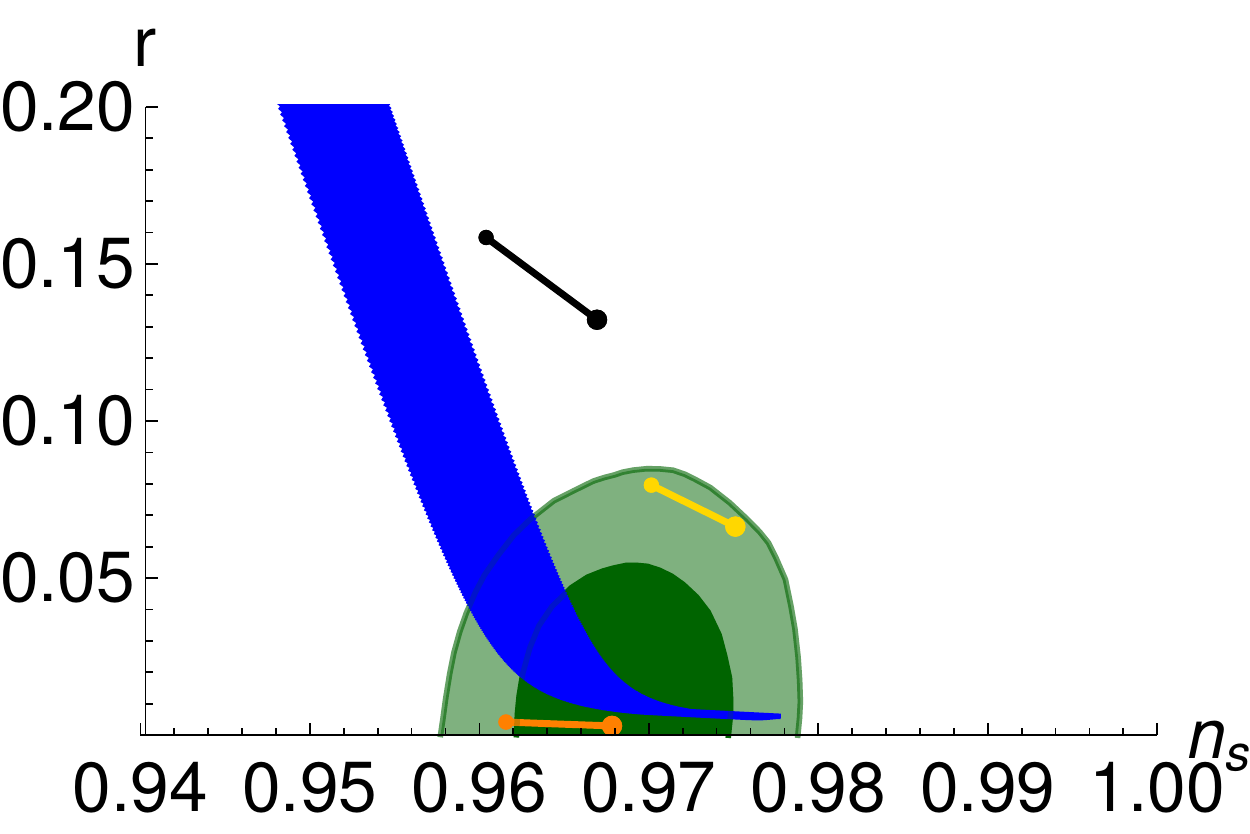}
  \includegraphics[width=0.49\textwidth]{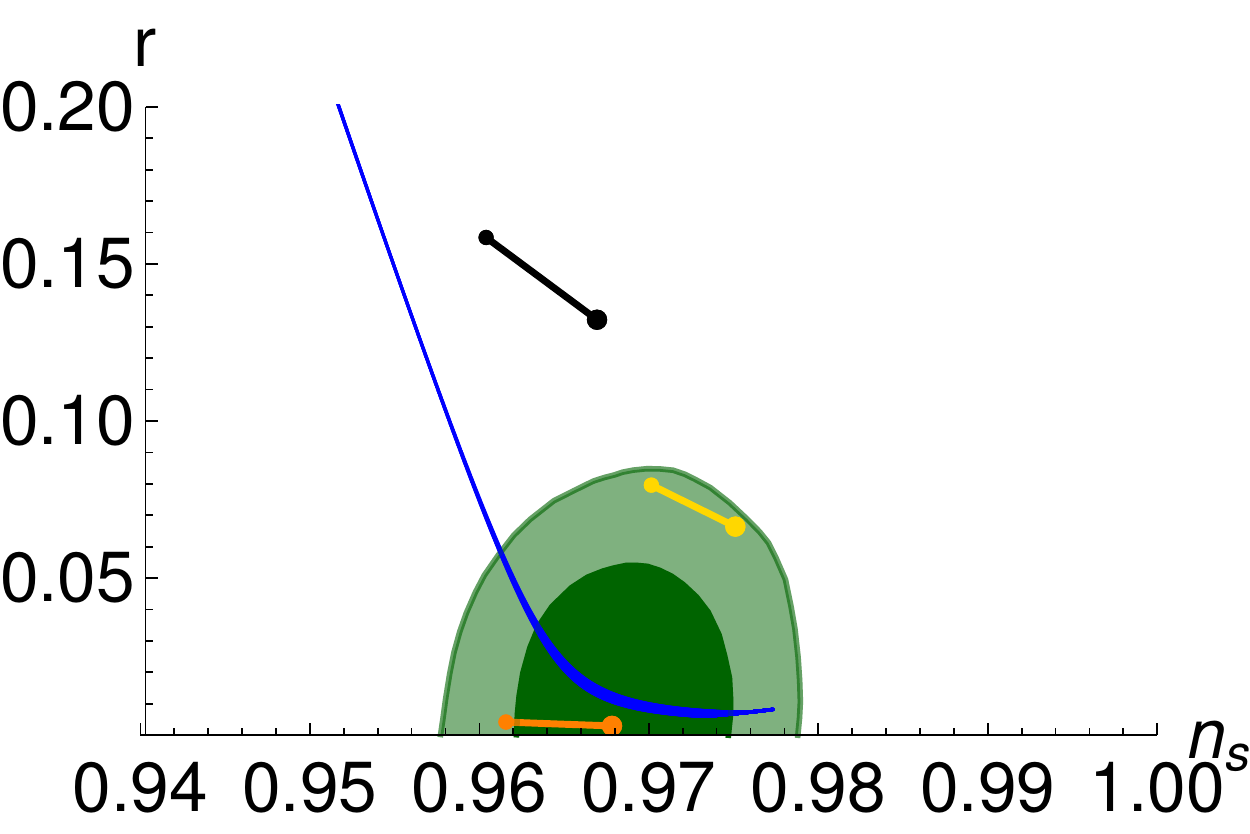}
  \caption{Inflationary results.  Dark and light green are the $1\sa$ and $2\sa$ regions from the new BICEP2/Keck Array data \cite{BICEP2new}.  Black/yellow/orange are quadratic/linear/Starobinsky inflation.  Blue points are from our model.  Left: $v_\phi=10^5$~GeV and $50<N_e<60$.  Right: $50$~GeV $\leq v_\phi \leq 10^8$~GeV and $N_e=55$. For numerical purposes we used an upper bound $\xi < e^{15}$.}
  \label{fig:inf}
\ec
\end{figure}

At this point we have all the pieces needed to assemble our model of inflation. The calculation proceeds henceforth as usual within slow-roll Higgs inflation models. Fig.~\ref{fig:inf} presents our results for the scalar index $n_s$ and the tensor-to-scalar ratio $r$. We plot the configuration $v_\phi=10^5$~GeV for $50<N_e<60$ in the left panel, while in the right panel we consider $50$~GeV $\leq v_\phi \leq 10^8$~GeV and $N_e=55$. For numerical purposes we used an upper bound $\xi < e^{15}$; the configurations that will satisfy the diphoton signal are anyhow inside such a bound. The blue points mark the region selected by our model while the dark and light green areas denote the BICEP2/Keck Array $1\sa$ and $2\sa$ confidence region \cite{BICEP2new}, respectively. The black/yellow/orange line shows the results of quadratic, linear and Starobinsky inflation. As we can see from Fig.~\ref{fig:inf}, for increasing values of $\xi_\phi$ we depart from the quartic inflationary configuration, decreasing the value of $r$ until reaching a sort of plateau around $r \approx 0.006-0.008$ (according to the exact value of $v_\phi$) and $n_s \approx 0.97$, which falls inside the $1\sa$ region of \cite{BICEP2new}.

In order to check whether our model can also explain the LHC diphoton excess, we scrutinise the successful inflation region --- the blue band in Fig.~\ref{fig:inf} --- looking for solutions where the inflaton acquires a mass $m_\phi$ around 750~GeV. Starting from the configuration $50$~GeV $\leq v_\phi \leq 10^8$~GeV and $N_e=55$, we required that the constraint on the amplitude of primordial scalar perturbations \cite{Ade:2015xua,Ade:2015lrj}
 \begin{equation}
A_s \pm \Delta A_s =(\PRexp)\times 10^{-9} \, , \label{eq:As}
 \end{equation}
be satisfied, and sought the solution yielding $m_\phi \approx 750$~GeV. Some resulting benchmark points satisfying both successful inflation and the LHC signal are reported in Tab.\ \ref{table:inf}.

\begin{table}[h!]
  \centering
\begin{tabular}{|c|c|c|c|c|c|}
  \hline			
  $v_\phi/$GeV & $\xi$ & $n_s$ & $r$& $m_\phi/$GeV \\
  \hline
100		&	$1.499 \times 10^6$	 &	0.9737	&	$6.178 \times 10^{-3}$	&	750.3	\\
500		&	$2.936 \times 10^5$	 &	0.9739	&	$6.261 \times 10^{-3}$	&	748.9	\\
$10^4$ 	& 	$1.419 \times 10^4$  & 	0.9742 	& 	$6.428 \times 10^{-3}$ 	& 	750.7\\
$10^6$ 	& 	133.0			     & 	0.9748 	& 	$6.729 \times 10^{-3}$ 	& 	747.9\\
$10^7$ 	& 	12.94 		         & 	0.9750 	& 	$6.932 \times 10^{-3}$ 	& 	754.0\\
 \hline
\end{tabular}
  \caption{A set of benchmark points that yield successful inflation and are compatible with the diphoton LHC signal. The points are obtained requiring  $N_e=55$ and 745~GeV $ < m_\phi < 755$~GeV.}
  \label{table:inf}
\end{table}

Looking at Tab.\ \ref{table:inf} we note that there is an inverse proportionality relation between the values of $\xi$ and $v_\phi$ which comply with the aforementioned requirements.  In order to further investigate this relation, we now temporarily move to the Einstein frame.  Given that $\xi \gg 1$, during inflation the relation between the canonically normalised fields in the two frames can be accurately approximated as
\be
  \phi \simeq \frac{M_\text{Pl}}{\sqrt\xi} e^{\frac{\phi_E}{\sqrt6 M_\text{Pl}}},
\ee
where $\phi_E$ is the inflaton field canonically normalised in the Einstein frame.  After performing the conformal transformation $g_{\mu\nu}\rar\Om^2g_{\mu\nu}$, the Einstein frame potential is
\bea
  V_E (\phi_E) &=& \frac{V(\phi(\phi_E))}{\Om^4(\phi(\phi_E))} \nn \\
&\simeq& \frac{M_\text{Pl}^4}{4\xi^2} \frac{m_{\phi}^2}{v_{\phi}^2} \left[\frac{\phi_E}{\sqrt6 M_\text{Pl}}+\ln \left(\frac{M_\text{Pl}}{v_\phi \sqrt\xi}\right)-\frac14\right] \left(1+e^{-\frac{\sqrt{\frac23} \phi_E}{M_\text{Pl}}}\right)^{-2} \, .
  \label{eq:VE}
\eea
The term isolated by round brackets is the usual Starobinsky term of Higgs inflation, while the term within square brackets is a new contribution from the running of $\la_\phi$.  Since $\ln x$ is slowly increasing for $x \gg 1$, the term $\ln \left(\frac{M_\text{Pl}}{v_\phi \sqrt\xi}\right)$ is roughly constant.  Then, in this approximation, only the overall normalisation of~(\ref{eq:VE}) depends on $v_\phi$ and $\xi$, implying that the scalar perturbations index and the tensor-to-scalar ratio $n_s$ and $r$ are independent of $v_\phi$ and $\xi$.
Moreover, by setting $m_\phi$ to the required value ($\approx$ 750 GeV), the overall normalisation of the potential is proportional to $1/\left(\xi v_\phi \right)^2$. As this combination is held constant by the constraint due to the amplitude of scalar perturbations, which sets the normalisation of the inflationary potential in this model, $v_\phi$ and $\xi$ must necessarily satisfy the inversely proportionality relation highlighted by our numerical results. Finally we also stress that our inflationary scenario is not strictly connected to an inflaton mass at 750 GeV. As clear from Eq. (\ref{eq:VE}), our model could fit all possible very light inflaton masses ($m_\phi \ll 10^{13}$ GeV), simply via a rescaling of the $\xi v_\phi$ product, so that the constraint from the amplitude of scalar perturbations remains satisfied.

\section{Inflaton decay rates and reheating}
\label{ssec:reh}

Reheating in this model proceeds through the scalar mediators $\sa_{i}$ into the SM.  A coupling between $\sa_{i}$ and the inflaton of the form
\begin{align}
	\mu_{i}\,\phi|\sa_{i}|^2
\end{align}
results in the following decay rates~\cite{Djouadi:2005gj}:
\begin{align}
	\label{eq:dwgaga0}
	\Ga_{\phi\ga\ga}
&	=	\frac{\alpha^{2}m_\phi^3}{1024\pi^{3}}
		\left|\sum_{i} \frac{d_{r_{i}}\mu_{i}}{m_{\sa_i}^{2}} q_{i}^{2} A_{0}\left(\frac{4m_{\sa_i}^{2}}{m_\phi^{2}}\right)\right|^{2},
	\\
	\label{eq:dwglgl0}
	\Ga_{\phi gg}
&	=	\frac{\alpha_{s}^{2}m_\phi^3}{8198\pi^{3}}
		\left|\sum_{i} \frac{d_{r_{i}}\mu_{i}}{m_{\sa_i}^{2}} C_{r_i} A_{0}\left(\frac{4m_{\sa_i}^{2}}{m_\phi^{2}}\right)\right|^{2},
\end{align}
where  $C_{r}$ and $d_{r}$ denote the first Casimir invariant and dimension of the colour representation $r$ (e.g. $C_{1} = 0$, $C_{3} = 4/3$ and $C_{8} = 3$), $q_{i}$ is the electric charge, and $\al$ and $\al_s$ are the electroweak and strong coupling constants, respectively. The scalar loop function reads
\begin{align}
	A_0(x) = x\left(1-x\,\arcsin^2(x^{-1/2})\right) \, ,
\end{align}
where the branch of the $\arcsin$ with positive imaginary values should be selected for $x < 1$. The sum is performed over all mediators.

For simplicity we assume $N_g$ mediators in a given colour representation and with a fixed electric charge $q$. Imposing an exact flavour symmetry on the mediator would imply $\mu_{i} = \mu$ and a degenerate spectrum $m_{\sigma_i} = m_{\sigma}$. Similar models have been studied in~\cite{Gabrielli:2015dhk,Knapen:2015dap}. The important difference in the current scenario is that here $m_\sigma$ and $\mu$ are not independent parameters because of our CW setup. The standard field redefinition $\phi \to \varphi + v_\phi$ gives $m_\sigma$ and $\mu$ as functions of the portal coupling $\lambda_{\phi\sigma}$ and the inflaton vev  $v_\phi$, where $m_\sigma$ is given in Eq.~\eqref{eq:msigma} and $\mu \equiv \lambda_{\phi\sigma} v_\phi$. Using (\ref{eq:mphi}, \ref{eq:beta_colour}) these relations can be rearranged yielding
\begin{equation}\label{eq:mu/mphi}
	\mu = \frac{4\pi}{\sqrt{\dof}}m_\phi .
\end{equation}
Given that the LHC diphoton signal fixes the inflaton mass at about 750 GeV, the decay rates (\ref{eq:dwgaga0},\ref{eq:dwglgl0}) depend only on $\lambda_{\phi\sigma}$, the charge $q$, and the eventual number of $\sigma$ copies considered. In conclusion, using Eq.~(\ref{eq:msigma}, \ref{eq:mu/mphi}) the decay widths can be recast as
\begin{align}
	\label{eq:dwgaga}
	\Ga_{\phi\ga\ga}
&	=	m_\phi \frac{\alpha^{2}\lambda_{\phi\sigma}^{2}q^{4}\dof^{3}}{8^{4}\pi^{5}}
	\left|A_{0}\left(\frac{32\pi^2}{\dof \lambda_{\phi\sigma}}\right)\right|^{2},
  \\	
  \label{eq:dwglgl}
	\Ga_{\phi gg}
&	=	m_\phi \frac{\alpha_{s}^{2}\lambda_{\phi\sigma}^{2}C_{r}^{2}\dof^{3}}{8^{5}\pi^{5}}
	\left|A_{0}\left(\frac{32\pi^2}{\dof \lambda_{\phi\sigma}}\right)\right|^{2}.
\end{align}

The requirement that the inflaton decay to the mediators is kinematically forbidden, $m_{\sa} > m_{\phi}/2$, implied by the LHC signal, yields a first upper bound on $\lambda_{\phi\sigma}$:
\begin{align}
	\lambda_{\phi\sigma} < \frac{32 \pi^{2}}{\dof}.
\end{align}
As remarked before, our CW setup requires that the couplings of the model remain well perturbative. In particular the condition $\lambda_{\phi\sigma} < 4 \pi$ must hold up to the scale of inflation, therefore we impose a rough but safe bound
\begin{align}
	\lambda_{\phi\sigma} \lesssim 1,
\end{align}
at the TeV scale. The first condition is stronger for $\dof > 315$.

Additionally, the constraints on $\lambda_{\phi\sigma}$ and $\left|A_{0}\right|^{2} < 2.15 $ in turn imply an upper bound for the partial decay widths
\begin{align}
	\label{bound:dwgaga}
	\Ga_{\phi \ga\ga}
&	\lesssim 7 \, \MeV \times q^{4} \times \dof \times \min(\dof/32\pi^{2},1)^{2},
	\\
	\label{bound:dwglgl}
	\Ga_{\phi gg}
&	\lesssim 130 \, \MeV \times C_{r}^{2} \times \dof \times \min(\dof/32\pi^{2},1)^{2},
\end{align}
where we used $m_{\phi} = 750 \GeV$ and $\alpha \simeq 1/137$, $\alpha_s(m_{\phi}/2) \simeq 0.09$.

Allowing the scalar mediator to carry both electric colour and charge generally results in $\Ga_\phi\equiv\Ga_{\phi gg}\gg\Ga_{\phi\ga\ga}$, in a way that only the former matters for the purpose of reheating.
As usual, we calculate the reheating temperature $T_r$ in the instantaneous reheating approximation
\be
  T_r^4 = \frac{90}{g_*\pi^2} \Ga_\phi^2 M_\text{Pl}^2 \, ,
\ee
where $g_*\approx100$ is the number of relativistic degrees of freedom in the Universe during the reheating epoch. Big Bang Nucleosynthesis (BBN) forces this temperature to be larger than a few MeV.  Within the range of values of the parameters suggested by LHC and Planck data it is easy to check that the (instantaneous) reheating temperatures are always large enough to accommodate for a fully successful BBN. The numerical results for reheating temperatures are presented in the last column of Table~\ref{table:LHC}.
The predicted reheating temperatures exceed ${\cal O}(10^8)$~GeV which also allow for successful leptogenesis.

\section{The diphoton excess}
\label{sec:dip}

In our picture, the inflaton is responsible for the excess observed in the diphoton channel of the ATLAS and CMS experiments~\cite{ATLAS2015, CMS2015}. In order to investigate the compatibility with the signal reported by these experiments, we estimate the cross section for the process $pp\to\phi\to\gamma\gamma$. By requiring that the latter falls in the measured ballpark of $5-10$ fb, we then constrain the inflaton vev as well as the charge and multiplicity of the scalar mediators.

In the narrow width approximation, the cross section for resonant production $pp \to \phi \to xx$ of a final state $xx$ via gluon-gluon fusion is
\begin{equation}\label{eq:sigma_pptogg}
	\sigma_{xx}
	=
	\frac{I_{gg}}{m_\phi^{3}}
	\frac{\Gamma_{\phi gg} \,\Gamma_{\phi xx}}{\Gamma_{\phi}}.
\end{equation}
where the dimensionless partonic integral
\begin{align}
	I_{gg}
	= \frac{\pi^2 }{8}\dfrac{m_\phi^{2}}{s}\int\limits_{m_\phi^2/s}^1
	\frac{\td x}{x} \, g(x) \, g\left(\frac{m\phi^2}{sx}\right)\,
	\approx 7.13,
\end{align}
was evaluated numerically at $\sqrt{s} = 13$ TeV using the MSTW parton distribution function (pdf) set \cite{Martin:2009iq}. Here, $g(x)$ denotes the pdf of the gluons evaluated at the scale 750 GeV. The partial decay widths to photons and gluons, that enter the calculation of the diphoton cross section,  are respectively given in Eq.~\eqref{eq:dwgaga} and Eq.~\eqref{eq:dwglgl}.

As before, when $\Ga_\phi\approx\Ga_{\phi gg}\gg\Ga_{\phi\ga\ga}$, the cross section~\eqref{eq:sigma_pptogg} simplifies, $\sigma_{\gamma\gamma} \simeq I_{gg}\Gamma_{\phi\gamma\gamma}m_\phi^{-3}$, and thus the perturbativity bound~\eqref{bound:dwgaga} on the partial decay width to photons implies a bound on the cross section $\sigma_{\gamma\gamma} \lesssim 4.5\times 10^{-4} \,{\rm fb} \times q^{4} \times \dof^{3}$. The observed cross-section of $5-10$ fb therefore requires a large number of scalar mediators or large electric charges. We remark that the approximation used to obtained the bound under consideration is not valid for values of electric charges much larger than one, since in that case the branching ratio to photons could exceed the branching ratio to gluons. The need for a large number of mediators (or non-perturbative couplings) is not a surprising result of our analysis, rather, it seems to be a generic feature of all perturbative models for the diphoton excess~\cite{Gabrielli:2015dhk,Knapen:2015dap}. Benchmark points for different models are given in Table \ref{table:LHC}.

\begin{table}[h!]
  \centering
\begin{tabular}{|c|c|c|c|c|c|c|c|c|}
  \hline			
  $d_{r}$ &
  $q$ &
  $N_{g}$ &
  $\lambda_{\phi\sigma}$ &
  $\sigma_{\gamma\gamma}^{13}/\fb$&
  $\sigma^8_{gg}/\pb$&
  $m_\sigma/\TeV$ &
  $v_\phi/\TeV$ &
  $T_r/\GeV$ \\
  \hline
 	3	&	1 	&	3	&	1		&	0.017 & $4.6 \times 10^{-5}$ & 2.2	&	3.1 & $3.1 \times 10^7$	\\
	3	&	1 	&	24	&	0.9		&	8.6	  & $2.4 \times 10^{-2}$ & 0.8 &	1.2 & $6.3 \times 10^8$	\\
	8	&	3 	&	3	&	0.7		&	8.4	  & $1.4 \times 10^{-3}$ & 1.6	&	2.7 & $2.1 \times 10^8$	\\
	15	&	2 	&	3	&	0.5		&	9.0	  & $3.8 \times 10^{-2}$ & 1.4 &	2.8 & $8.3 \times 10^8$	\\
	27	&	1 	&	3	&	0.7		&	7.5	  & 1.3	                 & 0.89 &	1.5 & $4.5 \times 10^9$	\\
	64	&	1 	&	1	&	1		&	7.7	  & 8.7	                 & 0.83 &	1.2 & $1.2 \times 10^{10}$	\\
  \hline
\end{tabular}
  \caption{A set of benchmark points obtained by setting $m_{\phi} = 750$ GeV and $\alpha \simeq 1/137$, $\alpha_s(m_{\phi}/2) \simeq 0.09$. The overall number of scalar mediators is $\dof = d_r N_g$. We indicate with $\sigma_{\gamma\gamma}^{13}$ and $\sigma^8_{gg}$ the diphoton and dijet cross sections respectively computed at 13 and 8 TeV of centre-of-mass energy.}
  \label{table:LHC}
\end{table}

On top of reproducing the detected diphoton signal, our scenario respects the constraints due to the complementary collider searches which target $WW$, $Z\gamma$, $ZZ$, dijet and monojet final states. In our case, given that the scalar mediators are singlets under the gauge group of weak interactions, we expect a vanishing branching ratio of the $750$ GeV inflaton into $WW$. Furthermore, the cross sections for the $Z\gamma$ and $ZZ$ channels automatically respect the corresponding experimental bounds \cite{Franceschini:2015kwy} being suppressed, respectively, by two and four powers of the sine of the Weinberg angle with respect to the diphoton one. As for the remaining final states, the monojet constraints can be evaded by simply requiring that the the scalar mediator be heavier than half the mass of the inflaton. Finally, we expect that searches for resonances in the dijet final state cast the strongest bounds on our model because of the large gluon decay that characterises it. In this regard, Eq.~\eqref{bound:dwglgl} provides a theoretical upper bound on the dijet cross section that holds in our framework
\begin{equation}
	\sigma_{gg} \approx \frac{I_{gg}}{m_\phi^3}\,\Gamma_{gg}
	\lesssim 6.96\times10^{-2}\text{ pb }\times C_r^2\times\dof\, ,
\end{equation}
where we used $I_{gg} \approx 0.58$ at 8 TeV centre-of-mass energy. The above condition is stronger than the one implied by the LHC dijet measurement, $\sigma_{gg}\lesssim 2.5$ pb at 8 TeV \cite{Franceschini:2015kwy}, for $C_r^2 \times \dof \lesssim 36$. Tab.~\ref{table:LHC} includes the values obtained for the dijet cross section in the corresponding benchmark points of the considered models: all but the last option are viable choices which satisfy all collider bounds and measurements, as well as cosmological requirements.

\section{Conclusions}
\label{sec:end}

If the LHC 750~GeV resonance is confirmed a new elementary singlet scalar field, a most natural interpretation would be to identify the new particle with the cosmological inflaton. Thence, under the assumption that the LHC has discovered nothing else but the inflaton, in this paper we analyse the compatibility of such a $750$ GeV inflaton with the LHC signal. We show that by barring explicit mass terms for the inflaton and for the SM Higgs boson in the Lagrangian, the interactions of the former with a secluded scalar field (implied by the detected diphoton excess) generate as well a Coleman-Weinberg potential suitable for inflation. Because of the portal coupling with the Higgs boson, the resulting inflaton potential is furthermore able to induce the SM electroweak breaking radiatively.
Therefore, in this scheme, the same particles that connect the inflaton to the SM and give rise to the LHC diphoton signal are also responsible for the generation of the inflationary potential.
Once a non-minimally coupling of the inflaton field to gravity is considered, we obtain a predicted tensor-to-scalar ratio of about 0.006, which allows to discriminate between our model and competing scenarios such as Starobinsky inflation, for instance. We find that the compatibility between our inflation and the detected LHC diphoton signal requires that the secluded scalar mediators have exotic properties: for instance, these particles might belong to large representations of the SM gauge groups or possess large electric charges.

\section*{Acknowledgments}
This work was supported by the ERC grants IUT23-6, PUTJD110, PUT 1026 and by the EU through the ERDF CoE program.

\bibliographystyle{JHEP}
\bibliography{citations}

\end{document}